\documentclass[prd,showpacs,preprintnumbers,amsmath,amssymb]{revtex4}

\usepackage{graphicx}
\usepackage{dcolumn}
\usepackage{bm}

\begin{document}

\title{Vacuum Persistence and Inversion of Spin Statistics in Strong QED}

\author{W-Y. Pauchy Hwang}\email{wyhwang@phys.ntu.edu.tw}
\affiliation{The Leung Center for Cosmology and Particle Astrophysics,
\\Institute of Astrophysics, Center for Theoretical Physics,
\\Department of Physics, National Taiwan University,
Taipei 106, Taiwan}

\author{Sang Pyo Kim}\email{sangkim@kunsan.ac.kr}
\affiliation{Department of Physics, Kunsan National University,
Kunsan 573-701, Korea,} \affiliation{Asia
Pacific Center for Theoretical Physics, Pohang 790-784, Korea}

\date{}

\begin{abstract}
The vacuum persistence can be written as the Bose-Einstein
distribution in spinor QED and as the Fermi-Dirac distribution in
scalar QED exactly in a constant electric field and approximately in
time-dependent electric fields. The inverse temperature is
determined by the period of charged particle in the Euclidean time
and the negative chemical potential by the ratio of the worldline
instanton to the inverse temperature. The negativity of chemical
potential is due to the vacuum instability under strong electric
fields. The inversion of spin statistics in the vacuum persistence
is a consequence of the Bogoliubov relations for fermions and
bosons.
\end{abstract}
\pacs{12.20.-m, 11.15.Tk, 13.40.-f, 12.20.Ds}

\maketitle

\section{Introduction}

The vacuum structure of quantum electrodynamics (QED) in external
electromagnetic fields exhibits many interesting features such as
vacuum polarization and pair production from an electric field due
to vacuum instability. The effective action of QED in a constant
electromagnetic field has been known by the seminal works by Sauter,
Heisenberg, and Euler, and Weisskopf in the 1930s \cite{Sauter} and
in a gauge invariant form via the proper-time method by Schwinger
\cite{Schwinger}. Physics of QED in strong electromagnetic fields
has recently attracted much attention partly because X-ray free
electron lasers \cite{Ringwald} and the intensive light source from
extreme light infrastructure (ELI) \cite{Dunne-Gies} may produce
electric fields near the critical strength $(E_c = 1.3 \times
10^{16}~{\rm V/cm})$ for electron-position pair production and
partly because neutron stars or magnetars are believed to have
fields ranging from $10^{8}~{\rm G}$ to $10^{15}~{\rm G}$
\cite{Wood-Thompson}, some of which go beyond the critical strength
$(B_c= 4.4 \times 10^{13}~{\rm G})$.

One of interesting aspects of QED effective action is the
observation by M\"{u}ller, Greiner and Rafelski \cite{MGR} that the
vacuum polarization, the real part of the effective action, in a
constant electric field reveals an inversion of spin statistics
between scalar and spinor QED. They showed that the vacuum
polarization of spinor QED can be written as a spectral function
times the Bose-Einstein distribution and that of scalar QED as the
same spectral function times the Fermi-Dirac distribution as
\begin{eqnarray}
{\rm Re} ({\cal L}_{\rm eff}) = - \frac{(2 |\sigma|+ 1) m^4}{16 \pi^2}
\int_{0}^{\infty} ds \Bigl[s \ln (s^2 - 1 + i \delta)
+ \ln \Bigl(\frac{s + 1 - i \delta}{s - 1 + i \delta}
\Bigr) -2s \Bigr] \frac{1}{e^{\beta s} + (-1)^{2 \sigma}},
\end{eqnarray}
where $\sigma = \pm 1/2$ for spinor QED and $\sigma = 0$ for scalar
QED and $\delta$ is an arbitrary small positive number and $\beta =
\pi m^2/qE$. One may show that the vacuum polarization in a pulsed
electric field of Sauter-type $E(t) = E_0 {\rm sech}^2 (t/\tau)$
\cite{Dunne-Hall,KLY} also exhibits the spin-statistics inversion
\cite{dunne-com}. Quite recently it has been shown that the WKB
instanton action for a pulsed electric field of Sauter-type
configuration in spinor QED more accurately yields the mean number
of produced pairs for scalar QED and vice versa
\cite{Kim-Page07,Kim09}. In quantum kinematic approach, the apparent
inversion of spin statistics is also shown for the Sauter-type
electric field \cite{Dumlu} and for an oscillating electric field
with the Gaussian envelope \cite{HADG}.

The purpose of this paper is to investigate the inversion of spin
statistics in the vacuum persistence, twice the imaginary part of
the effective action, both in a constant electric field and in a
general electric field. We show that the vacuum persistence of
spinor QED takes the Bose-Einstein distribution and that of scalar
QED the Fermi-Dirac distribution exactly in a constant electric
field and approximately in a general electric field. In the case of
a constant electric field, the temperature is determined by the
acceleration of a charged particle in analogy of the Hawking-Unruh
effect, as expected from Ref. \cite{MGR}. It is shown that the rest
mass energy provides a negative chemical potential to the
distributions, which implies the instability of the Dirac sea. For a
general electric field, the temperature and the chemical potential
for distributions depend not only on the mass and charge but also on
the electric profile, for instance, for a Sauter-type electric
field, $E(t) = E_0 {\rm sech}^2 (t/\tau)$, they are determined by
three parameters: the mass $m$, the field strength $E_0$ and the
duration $\tau$. This result together with the inversion of spin
statistics in the vacuum polarization may shed light on
understanding the vacuum structure of QED.

The organization of this paper is as follows. In Sec. II, we review
the general relation between the vacuum persistence and the mean
number of produced pairs. In Sec. III, we express the vacuum
persistence exactly in terms of the Bose-Einstein or Fermi-Dirac
distribution in a constant electric field and approximately in
time-dependent electric fields, including the Sauter-type electric
field. We discuss the temperature and chemical potential from the
dynamics of charged particle in electric fields in the Euclidean
time. In Sec. IV, we discuss and conclude the inversion of spin
statistics.

\section{Vacuum Persistence and Pair Production Rate}

The Heisenberg-Euler effective Lagrangian in a constant
electromagnetic field has been obtained using various methods (for
references and review, see Ref. \cite{Dunne}). One of the methods is
the evolution operator method that directly makes use of the
Bogoliubov transformation, which transforms the in-vacuum to the
out-vacuum \cite{KLY}. It is based on the idea that the scattering
amplitude between the in-vacuum and the out-vacuum leads to the
effective action \cite{Schwinger51}. For temporal evolution, the
out-vacuum operators are expressed in terms of the in-vacuum
operators through the Bogoliubov transformation, which enables the
effective action to be written as the Bogoliubov coefficients
\cite{Nikishov70,DeWitt,AHN,Nikishov03,GGT}. In this section, we
shall employ the evolution operator method in Ref. \cite{KLY} to
compute the QED effective action in time-dependent electric fields
and the mean number of produced pairs.

In the evolution operator method, we use the Bogoliubov
transformation between the in-vacuum and the out-vacuum given by
\begin{eqnarray}
\hat{a}^{\rm out}_{{\bf k} \sigma} = \mu_{ {\bf
k} \sigma} \hat{a}^{\rm in}_{ {\bf k} \sigma}
+ \nu^*_{ {\bf k} \sigma} \hat{b}^{{\rm
in}\dagger}_{ {\bf k} \sigma},
\end{eqnarray}
where $\hat{a}$ and $\hat{b}$ denote the particle and antiparticle
operators, respectively. Here, the Bogoliubov coefficients satisfy
the relation
\begin{eqnarray}
| \mu_{{\bf k} \sigma} |^2 - (-1)^{2 \sigma}  |
\nu_{{\bf k} \sigma}|^2 = 1.
\end{eqnarray}
Remarkably the Bogoliubov transformation both in spinor QED and in
scalar QED can be written through the evolution operator as
\begin{eqnarray}
\hat{a}^{\rm out}_{{\bf k} \sigma} = \hat{\rm U}_{{\bf k} \sigma} (\hat{a}^{\rm in}_{{\bf k} \sigma}, \hat{b}^{\rm in}_{{\bf k} \sigma}) \hat{a}^{\rm in}_{{\bf k} \sigma} \hat{\rm U}^{\dagger}_{{\bf k} \sigma} (\hat{a}^{\rm in}_{{\bf k} \sigma}, \hat{b}^{\rm in}_{{\bf k} \sigma}), \label{bog-ev}
\end{eqnarray}
where the explicit form of $\hat{\rm U}$ is given in Ref.
\cite{KLY}. In particular, Eq. (\ref{bog-ev}) implies that the
out-vacuum evolves from the in-vacuum as
\begin{eqnarray}
\vert 0, {\rm out} \rangle = \prod_{{\bf k} \sigma}
\hat{\rm U}_{{\bf k} \sigma} \vert 0, {\rm in} \rangle.
\end{eqnarray}
Finally, the effective action defined as the scattering amplitude
takes the form \cite{KLY}
\begin{eqnarray}
e^{i \int dt d^3x {\cal L}_{\rm eff}} = \langle 0, {\rm out} \vert 0, {\rm in} \rangle = e^{-(-1)^{2 \sigma} \sum_{{\bf k} \sigma} \mu^{*}_{ {\bf
k} \sigma}}.
\end{eqnarray}
Thus, the one-loop effective action per unit time and per unit volume is
\begin{eqnarray}
{\cal L}_{\rm eff} = (-1)^{2 \sigma} i \sum_{ {\bf
k} \sigma} \ln (\mu^*_{ {\bf k} \sigma}),
\label{eff act}
\end{eqnarray}
which leads to the general relation between the vacuum persistence
(twice of the imaginary part of the effective action) and the mean
number of produced pairs, ${\cal N}_{{\bf k} \sigma} = |\nu_{{\bf k}
\sigma}|^2$:
\begin{eqnarray}
2 ({\rm Im} {\cal L}_{\rm eff}) = (-1)^{2 \sigma} \sum_{{\bf k}
\sigma} \ln [ 1 + (-1)^{2 \sigma} {\cal N}_{{\bf
k} \sigma}].
\end{eqnarray}

\section{Inversion of Spin Statistics}

In this section we shall study the vacuum persistence exactly in a
constant electric field and approximately in time-dependent electric
fields, in particular, a pulsed electric field of Sauter-type.

First, we consider the exact case of a constant electric field as in
Ref. \cite{MGR}. In the time-dependent gauge, the Klein-Gordon
equation for scalar QED and the Dirac equation for spinor QED have
the (spin diagonal) Fourier component [in natural units with $\hbar
= c = 1$]
\begin{eqnarray}
\Bigl[\partial_t^2 + (k_z - qEt)^2 + 2m \Bigl(\epsilon+ \frac{m}{2}
\Bigr) + 2 i \sigma q E \Bigr] \phi_{{\bf k} \sigma} (t) = 0,
\label{time-comp}
\end{eqnarray}
where $\sigma = \pm 1/2$ for spinor QED and $\sigma = 0$ for scalar
QED and $\epsilon$ is the excitation energy in the transverse
direction in unit of the particle mass:
\begin{eqnarray}
\epsilon = \frac{{\bf k}_{\perp}^2}{2m}.
\end{eqnarray}
The solution to Eq. (\ref{time-comp}) given by the parabolic
cylinder function
\begin{eqnarray}
\phi_{{\bf k} \sigma} (t) = D_p(\zeta),
\end{eqnarray}
where
\begin{eqnarray}
\zeta = \sqrt{2qE} e^{i \pi/4} \Bigl( \frac{k_z}{qE} - t \Bigr),
\quad p = - \Bigl(\frac{1}{2} + \sigma \Bigr) - i \frac{m( \epsilon+
\frac{m}{2})}{qE},
\end{eqnarray}
defines asymptotically the in-vacuum at $t = - \infty$ and the
out-vacuum at $t = \infty$. Then the Bogoliubov coefficients are
\begin{eqnarray}
\mu_{{\bf k}_{\perp} \sigma} = \frac{\sqrt{2 \pi}}{\Gamma(-p)}
e^{-i (p+1)\pi/2}, \quad \nu_{{\bf k}_{\perp} \sigma} = e^{-i p \pi}.
\end{eqnarray}
Following the renormalization scheme in Ref. \cite{KLY}, the vacuum
polarization is then given by
\begin{eqnarray}
{\rm Re} ({\cal L}_{\rm eff}) = (-1)^{2 |\sigma| + 1} \frac{(2 |\sigma| +
1)}{16 \pi^2} {\cal P} \int_0^{\infty} \frac{ds}{s^3} e^{ - m^2 s} [
(qEs) f(s) - g(s)], \label{spin-E3}
\end{eqnarray}
where ${\cal P}$ denotes the principal value, and $f(s)= {\rm
cosec}(qEs)$, $ g(s) = 1 + (qE s)^{2}/6$ for scalar QED and $f(s) =
\cot(qE s)$, $g(s) = 1/s - (qES)^3/3$ for spinor QED. On the other
hand, the vacuum persistence takes the form
\begin{eqnarray}
2 {\rm Im} ({\cal L}_{\rm eff}) = - (-1)^{2 \sigma} \frac{(2
|\sigma| + 1) qE}{2\pi} \int \frac{d^2 {\bf k}_{\perp}}{(2 \pi)^2}
\sum_{n =1}^{\infty}  \Bigl[ - (-1)^{2 \sigma} e^{- \frac{2 \pi
m(\epsilon+ \frac{m}{2})}{qE}} \Bigr]^n, \label{multi-instanton}
\end{eqnarray}
where, in the last factor, the summation over residues from poles
along the imaginary axis in the first quadrant has an interpretation
of multi-instantons and anti-instantons \cite{Kim-Page02}. Summing
over $n$ and changing the variable $d^2 {\bf k}_{\perp} / (2\pi)^2 =
m d \epsilon/2\pi$, we may write the vacuum persistence as
\begin{eqnarray}
2 {\rm Im} ({\cal L}_{\rm eff}) = (-1)^{2 \sigma} \frac{(2 |\sigma|
+ 1) qE m}{4 \pi^2} \int_{0}^{\infty}  d \epsilon \ln [1 + (-1)^{2
\sigma} {\cal N}_{\epsilon}], \label{const-E}
\end{eqnarray}
where ${\cal N}_{\epsilon}$ is the mean number of produced pairs:
\begin{eqnarray}
{\cal N}_{\epsilon} = e^{- \frac{2 \pi m( \epsilon +
\frac{m}{2})}{qE}}.
\end{eqnarray}

We now wish to give a thermal interpretation of the vacuum
persistence (\ref{const-E}). The physical intuition of temperature
for a charged particle or antiparticle in a constant electric field
is the Hawking-Unruh temperature \cite{Hawking,Unruh}
\begin{eqnarray}
T_{\rm HU} = \frac{a}{2 \pi},
\end{eqnarray}
associated with the acceleration $a$ or the surface gravity $a$ of a
black hole. An accelerated observer measures the thermal state with
the Hawking-Unruh temperature from the Minkowski vacuum due to the
presence of a horizon. In fact, the acceleration of a charged
particle in a constant electric field
\begin{eqnarray}
a = \frac{qE}{m}
\end{eqnarray}
leads to the inverse Hawking-Unruh temperature \cite{Stephens}
\begin{eqnarray}
\beta_0 = \frac{1}{k_{\rm B} T_{\rm HU}} = \frac{2\pi m}{qE}.
\end{eqnarray}
Another interpretation of temperature is the acceleration of the
reduced mass, $m/2$, of a particle and antiparticle pair
\cite{CKS07}. Integrating in parts, we may write the vacuum
persistence as
\begin{eqnarray}
2 {\rm Im} ({\cal L}_{\rm eff}) = \frac{(2 |\sigma| + 1)m^2}{2\pi}
\int_{0}^{\infty} d \epsilon \frac{\epsilon}{ e^{\beta_0 ( \epsilon
+ \frac{m}{2})} + (-1)^{2 \sigma}}. \label{vac-dist}
\end{eqnarray}
A few comments are in order. First, note that Eq. (\ref{vac-dist})
describes the Bose-Einstein distribution for spin-1/2 fermions,
whereas it does the Fermi-Dirac distribution for spin-0 scalars.
Second, the negative chemical potential $\eta_0 = -m/2$
\cite{Stephens} in natural units implies that pair production is
more favored because the vacuum is unstable against pair production
in the presence of strong electric fields. Finally, the connection
between the instanton action in Eq. (\ref{multi-instanton}) and the
Matsubara frequency may be physically feasible since the
Bose-Einstein or Fermi-Dirac distribution in Eq. (\ref{vac-dist})
always has an expression in terms of the Matsubara frequency in
finite-temperature field theory.

In the second case of a general time-dependent electric field with
the gauge potential, $A_3 (t) = - E_0 f(t)$, the Klein-Gordon
equation and the spin diagonal component of the Dirac equation take
the form
\begin{eqnarray}
[ \partial_t^2 +  Q_{ {\bf k} \sigma} (t)] \phi_{{\bf
k} \sigma} (t) = 0, \label{gen eq}
\end{eqnarray}
where
\begin{eqnarray}
Q_{{\bf k} \sigma}(t) = (k_z - qE_0 f(t))^2 + m \Bigl(\epsilon +
\frac{m}{2} \Bigr) + 2 i \sigma q E_0 \dot{f}(t).
\end{eqnarray}
Though the exact solution of Eq. (\ref{gen eq}) may not be found in
general, the WKB method can provide a scheme to find approximately
the mean number of produced pairs per unit volume and per momentum,
which is given by \cite{Kim-Page07}
\begin{eqnarray}
{\cal N} (\epsilon, k_z, \sigma) = e^{- {\cal S} (\epsilon, k_z, \sigma)}, \quad {\cal S} (\epsilon, k_z, \sigma) = i \oint \sqrt{Q_{{\bf k} \sigma}(t)}dt.
\end{eqnarray}
Then, the vacuum persistence can be written as
\begin{eqnarray}
2 {\rm Im} ({\cal L}_{\rm eff}) = \frac{m}{2 \pi} \sum_{\sigma}
\int_{- \infty}^{\infty} \frac{d k_z}{2 \pi} \int_{0}^{\infty} d
\epsilon \frac{\epsilon d {\cal S}(\epsilon, k_z, \sigma)/d
\epsilon}{ e^{{\cal S}(\epsilon, k_z, \sigma)} + (-1)^{2 \sigma}}.
\end{eqnarray}

For instance, the Sauter-type electric field \cite{Sauter32}, $E(t)
= E_0 {\rm sech}^2 (t/\tau)$ with $f(t) = \tau \tanh (t/\tau)$, has
the WKB action \cite{Kim-Page07}
\begin{eqnarray}
{\cal S} (\epsilon, k_z, \sigma) &=& \pi qE_0 \tau^2
\Biggl[\sqrt{\Bigl(1+ \frac{k_z}{qE_0 \tau} \Bigr)^2 + \frac{2m (
\epsilon + \frac{m}{2})}{(qE_0 \tau)^2}} \nonumber\\&& +
\sqrt{\Bigl(1- \frac{k_z}{qE_0 \tau} \Bigr)^2 + \frac{2m(\epsilon
+\frac{m}{2} )}{(qE_0 \tau)^2}} - \lambda \Biggr], \label{WKB
action}
\end{eqnarray}
where $\lambda = 2$ for spinor QED and $\lambda = \sqrt{1-1/(2qE_0
\tau^2)^2}$ for scalar QED. The WKB action up to quadratic order of
the momentum is equivalent to the worldline instanton with the
prefactor included \cite{Dunne-Schubert}. In the region where the
WKB action and the worldline instanton are a good approximation,
$\lambda =2$ approximately even for scalar QED and we shall not
distinguish spinor QED from scalar QED. Expanding up to quadratic
order of momentum and changing the variable $k_z = (qE_0 \tau\sqrt{1
+ (m/qE_0 \tau)^2}) \omega$, we obtain approximately the vacuum
persistence per unit volume,
\begin{eqnarray}
2 {\rm Im} ({\cal L}_{\rm eff}) \approx \frac{(2 |\sigma| +1) m^2
\tau}{2 \pi} \int_{- \infty}^{\infty} d \omega \int_{0}^{\infty} d
\epsilon \frac{\epsilon}{ e^{\beta (\epsilon + \frac{m}{2} \omega^2
- \eta)} + (-1)^{2 \sigma}}, \label{sauter}
\end{eqnarray}
where the temperature $\beta$ and the chemical potential $\eta$ are
\begin{eqnarray}
\beta = \frac{2 \pi m}{qE_0\sqrt{1 + (\frac{m}{qE_0\tau}})^2}, \quad
\eta = - m \frac{\sqrt{1+ (\frac{m}{qE_0 \tau})^2}}{1+ \sqrt{1+
(\frac{m}{qE_0 \tau})^2}}.
\end{eqnarray}
In the limit of $\tau = \infty$, corresponding to a constant
electric field, $\omega =0$ and the $\omega$-integration becomes
unity, which makes Eq. (\ref{sauter}) reduce to the vacuum
persistence (\ref{vac-dist}) per unit volume and per unit time in
the constant electric field, as expected. Further, the temperature
and the chemical potential become $\beta_0 = 2 \pi m/qE$ and $\eta_0
= -m/2$, respectively. In Ref. \cite{Chervyakov-Kleinert}, the
pair-production rate (95) in scalar QED in a spatially localized
electric field of Sauter-type, which is the vacuum persistence of
this paper, is written as a spectral function times the Fermi-Dirac
distribution, in concord with this paper.

Finally, we associate the dynamics of a charged particle or
antiparticle in electric fields with the temperature and the
chemical potential of distributions. In the Euclidean time, the
particle follows a closed trajectory \cite{Dunne-Schubert}
\begin{eqnarray}
\dot{x}^2_3 (u) + \dot{x}^2_4 (u) = c^2.
\end{eqnarray}
Here, $u$ is the Euclidean time and $c$ is a constant related with
the mass, charge and field configuration. The particle executes a
circular motion in the constant electric field, whose period is
$\beta_0$ for $c=1$, whereas it follows a complicated closed path in
the Sauter-type electric field, whose period is $\beta$. In both
electric fields, the chemical potential is $\eta = - S_0/\beta$,
where $S_0$ is the worldline instanton \cite{Dunne-Schubert}
\begin{eqnarray}
S_0 = \frac{\pi m^2}{qE}\frac{2}{1+ \sqrt{1+ (\frac{m}{qE_0 \tau})^2}}.
\end{eqnarray}
The worldline instanton $S_0$ is the zero-momentum limit of the WKB
action ${\cal S} (\epsilon = 0, k_z = 0)$ in Eq. (\ref{WKB action}).

\section{Conclusion}

M\"{u}ller, Greiner, and Rafelski \cite{MGR} showed that the vacuum
polarization of spinor QED, the real part of the effective action,
in a constant electric field can be expressed as a spectral function
times the Bose-Einstein distribution and that of scalar QED as the
same spectral function times the Fermi-Dirac distribution, thus
exhibiting the spin-statistics inversion. In this paper, we showed
that the vacuum persistence, twice of the imaginary part of the
effective action, also exhibits the inversion of spin statistics
between spinor and scalar QED exactly in the constant electric field
and approximately in time-dependent electric fields. The inversion
of spin statistics is a consequence of the Bogoliubov relations for
fermions and bosons, which are equivalent to equal-time commutation
relations for fermions and bosons that are closely related with spin
statistics.

The temperature and the chemical potential for distributions are
determined by the dynamics of charged particle in electric fields.
In the constant electric field, the period of a circular motion in
the Euclidean time is the inverse temperature and the ratio of the
worldline instanton or the zero-momentum of the WKB action to the
inverse temperature is the chemical potential. A similar argument
holds for the Sauter-type electric field, though the motion is
complicated. Another interpretation is the Hawking-Unruh temperature
associated with the acceleration of particle in the constant
electric field. Though it is not an easy task to find the
Hawking-Unruh temperature for general time-dependent electric
fields, the dynamical approach based on the worldline instanton or
the WKB action does provide the temperature for the distribution in
the vacuum persistence.

The inversion of spin statistics of the vacuum polarization has been
known for a constant electric field. It would be interesting to see
whether such inversion of the vacuum polarization still works for
the Sauter-type electric field, whose vacuum persistence exhibits
approximately the inversion as shown in this paper.

\acknowledgments We would like to thank Professor J.~Rafelski for
informing the spin-statistics inversion of the QED effective action
in a constant electric field, S.~G.~Kim, H.~K.~Lee, and Y.~Yoon for
helpful discussions and G. Dunne for useful comments and informing
us of Ref. \cite{Stephens}. S.~P.~K. appreciates the warm
hospitality of National Taiwan University during the Symposium on
Symmetries of Subatomic Particle Physics 2009 (SSP2009). The work of
S.~P.~K. was supported by the Korea Research Foundation Grant funded
by the Korean Government (MOEHRD) (KRF-2007-313-C00167) and the
travel for SSP2009 was supported by the Korea Research Council of
Fundamental Science and Technology (KRCF). The work of W-Y.~P.~H.
was supported by the Leung Center for Cosmology and Particle
Astrophysics (LeCosPA), National Taiwan University.


\begin{thebibliography}{9}

\bibitem{Sauter} F.~Sauter, Z. Phys. {\bf 69}, 742
(1931); W.~Heisenberg and H.~Euler, Z. Physik {\bf 98}, 714 (1936);
V.~Weisskopf, Kgl. Danske Videnskab. Selskabs. Mat.-fys. Medd. 14,
No. 6 (1936).

\bibitem{Schwinger} J.~Schwinger, Phys. Rev. {\bf 82}, 664 (1951).

\bibitem{Ringwald} A. Ringwald, Phys. Lett. B {\bf 510}, 107 (2001);
``Fundamental Physics at an X-Ray Free Electron Laser,'' {\it
Electromagnetic Probes of Fundamental Physics}, edited by
W.~Marciano and S.~White (World Scientific, Singapore, 2003), pp.
63-74, hep-ph/0112254; ``Boiling the Vacuum with an X-Ray Free
Electron Laser,'' {\it Quantum Aspects of Beam Physics 2003}, edited
by P.~Chen and K.~Reil (World Scientific, Singapore, 2004), pp.
149-163, hep-ph/0304139.

\bibitem{Dunne-Gies}
H.~Gies, Eur. Phys. J. D, DOI: 10.1140/epjd/e2009-00006-0 (2009)
[arXiv:0812.0668];\\G.~V.~Dunne, Eur. Phys. J. D, DOI:
10.1140/epjd/e2009-00022-0 (2009) [arXiv:0812.3163].

\bibitem{Wood-Thompson} P.~M.~Wood and C.~Thompson, in {\it Compact
Stellar X-ray Sources}, edited by W.~H.~G.~Lewin and
M.~van~der~Klis (Cambridge University, Cambridge, 2006).

\bibitem{MGR} B.~M\"{u}ller, W.~Greiner and J.~Rafelski, Phys. Lett. A {\bf 63}, 181 (1977).


\bibitem{Dunne-Hall} G.~Dunne and T.~Hall, Phys. Rev. D {\bf 58},
105022 (1998).

\bibitem{KLY} S.~P.~Kim, H.~K.~Lee, and Y.~Yoon, Phys. Rev D {\bf 78},
105013 (2008); S.~P.~Kim and H.~K.~Lee, J. Korean Phys. Soc. {\bf 54}, 2605 (2009) [arXiv: 0806.2496].

\bibitem{dunne-com} Private communication with G. Dunne.

\bibitem{Kim-Page07} S.~P.~Kim and D.~N.~Page, Phys. Rev. D
{\bf 75}, 045013 (2007).

\bibitem{Kim09} S.~P.~Kim, AIP Conf. Proc. {\bf 1150}, 95 (2009) [arXiv:0902.3486].

\bibitem{Dumlu} C.~K.~Dumlu, Phys. Rev. D {\bf 79}, 065027 (2009).

\bibitem{HADG} F.~Hebenstreit, R.~Alkofer, G.~V.~Dunne, and H.~Gies,
Phys. Rev. Lett. {\bf 102}, 150404 (2009).


\bibitem{Dunne} G.~V.~Dunne, ``Heisenberg-Euler Effective Lagrangians: Basics
and Extensions,'' in {\it From Fields to Strings: Circumnavigating
Theoretical Physics}, edited by M.~Shifman, A. Vainshtein, and J.
Wheater, World Scientific, Singapore, 2005, Vol. I, pp. 445--522.

\bibitem{Schwinger51} J.~Schwinger, Proc. Natl Acad. Sci. U.S.A. {\bf 37}, 452 (1951).

\bibitem{DeWitt} B.~S.~DeWitt, Phys. Rep. {\bf 19}, 295 (1975).


\bibitem{AHN} J.~Ambjorn, R.~J.~Hughes and N.~K.~Nielsen, Ann. Phys. (N.Y.) {\bf 150}, 92 (1983).

\bibitem{Nikishov03} A.~I.~Nikishov, JETP {\bf 96}, 180 (2003).

\bibitem{Nikishov70} A.~I.~Nikishov, Sov. Phys. JETP {\bf 30}, 660 (1970) [Zh. Eksp. Teor. Fiz.
{\bf 57}, 1210 (1969)].

\bibitem{GGT} S.~P.~Gavrilov, D.~M.~Gitman and J.~L.~Tomazelli, Nucl Phys. {\bf B795}, 645 (2008).

\bibitem{Kim-Page02} S.~P.~Kim and D.~N.~Page, Phys. Rev. D {\bf 65}, 105002
(2002).

\bibitem{Hawking} S.~W.~Hawking, Comm. Math. Phys. {\bf 43}, 199 (1975).

\bibitem{Unruh} W.~G.~Unruh, Phys. Rev. D {\bf 14}, 870 (1976).

\bibitem{Stephens} C.~R.~Stephens, Ann. Phys. {\bf 193}, 255 (1989).

\bibitem{CKS07} P.~Castorina, D.~Kharzeev, and H.~Satz, Eur. Phys. J. C {\bf 52}, 187 (2007).

\bibitem{Sauter32} F.~Sauter, Z. Phys. {\bf 73}, 547 (1932).

\bibitem{Dunne-Schubert} G.~V.~Dunne and C.~Schubert, Phys. Rev. D
{\bf 72}, 105004 (2005); G.~V.~Dunne, Q.-H.~Wang, H.~Gies, and C.~Schubert, Phys. Rev. D {\bf 73}, 065028 (2006).

\bibitem{Chervyakov-Kleinert} A.~Chervyakov and H.~Kleinert, ``Exact Pair Production Rate for a Smooth Potential Step,'' [arXiv:0906.1422].


\end{thebibliography}
\end{document}